\documentclass[reprint,aps,prd,twocolumn,superscriptaddress,showpacs,preprintnumbers,amsmath,amssymb,floatfix,nofootinbib]{revtex4-2}

\usepackage[utf8]{inputenc}
\usepackage[export]{adjustbox} 
\usepackage{graphicx,amssymb,amsmath,hyperref,xspace,tabularx,pifont,floatrow}
\usepackage[caption=false]{subfig}

\usepackage{textcase}

\usepackage[capitalize]{cleveref}
\usepackage[normalem]{ulem} 

\usepackage{xcolor}

\definecolor{newyellow}{HTML}{FFEF00}

\begin{document}

\title{A Search for Low-mass Dark Matter via Bremsstrahlung Radiation and the Migdal Effect in SuperCDMS}


\author{M.F.~Albakry} \affiliation{Department of Physics \& Astronomy, University of British Columbia, Vancouver, BC V6T 1Z1, Canada}\affiliation{TRIUMF, Vancouver, BC V6T 2A3, Canada}
\author{I.~Alkhatib} \affiliation{Department of Physics, University of Toronto, Toronto, ON M5S 1A7, Canada}
\author{D.~Alonso} \affiliation{Instituto de F\'{\i}sica Te\'orica UAM/CSIC, Universidad Aut\'onoma de Madrid, 28049 Madrid, Spain}\affiliation{Instituto de F\'{\i}sica Te\'orica UAM-CSIC, Campus de Cantoblanco, 28049 Madrid, Spain}
\author{D.W.P.~Amaral} \affiliation{Department of Physics, Durham University, Durham DH1 3LE, UK}
\author{T.~Aralis} \affiliation{Division of Physics, Mathematics, \& Astronomy, California Institute of Technology, Pasadena, CA 91125, USA}
\author{T.~Aramaki} \affiliation{Department of Physics, Northeastern University, 360 Huntington Avenue, Boston, MA 02115, USA}
\author{I.J.~Arnquist} \affiliation{Pacific Northwest National Laboratory, Richland, WA 99352, USA}
\author{I.~Ataee~Langroudy} \affiliation{Department of Physics and Astronomy, and the Mitchell Institute for Fundamental Physics and Astronomy, Texas A\&M University, College Station, TX 77843, USA}
\author{E.~Azadbakht} \affiliation{Department of Physics and Astronomy, and the Mitchell Institute for Fundamental Physics and Astronomy, Texas A\&M University, College Station, TX 77843, USA}
\author{S.~Banik} \affiliation{School of Physical Sciences, National Institute of Science Education and Research, HBNI, Jatni - 752050, India}
% ...

\date{\today}%

\begin{abstract}
\clearpage

We present a new analysis of previously published of SuperCDMS data using a profile likelihood framework to search for sub-GeV dark matter (DM) particles through two inelastic scattering channels: bremsstrahlung radiation and the Migdal effect.
By considering these possible inelastic scattering channels, experimental sensitivity can be extended to DM masses that are undetectable through the DM-nucleon elastic scattering channel, given the energy threshold of current experiments.
We exclude DM masses down to $220~\textrm{MeV}/c^2$ at $2.7 \times 10^{-30}~\textrm{cm}^2$ via the bremsstrahlung channel. The Migdal channel search provides overall considerably more stringent limits and excludes DM masses down to $30~\textrm{MeV}/c^2$ at $5.0 \times 10^{-30}~\textrm{cm}^2$.

\end{abstract}

\collaboration{SuperCDMS Collaboration}
\noaffiliation

\maketitle


\section{Introduction}

An abundance of evidence suggests that most of the Universe is composed of non-luminous matter~\cite{PDG,Planck,Bullet_Cluster}. This ``dark matter” (DM) may consist of an undiscovered elementary particle or a set of particles~\cite{DM_history}. However, since particle DM has not been detected directly, its exact properties, such as mass and interaction cross section with standard model particles, have yet to be determined.

Much effort has been focused on searches for particles with masses in the GeV$/c^2$ to TeV$/c^2$ range, where the favored detection mechanism is rare collisions observed by terrestrial detectors~\cite{Cosmic_Visions}. Some of these approaches can be extended to reach below $1~\textrm{GeV}/c^2$ through inelastic detection channels. 
In canonical direct DM searches, the interaction between a DM particle and a nucleus is assumed to be an elastic two-body interaction. For DM particles with masses, $m_\chi$, much smaller than that of the target nucleus $m_N$, the recoil energy from an elastic collision is suppressed by the kinematic term $m_\chi^2/m_N$, resulting in rapidly diminishing sensitivity when considering lower mass DM candidates.
This suppression is the result of momentum and energy conservation with a heavy target nucleus, but it can be circumvented by involving a third particle in the scattering process when $m_\chi << m_N$.
In such inelastic scatterings, the third particle can receive up to the full energy of the collision~\cite{Brem_KP}. 
Detection of this higher energy particle provides sensitivity to DM masses that were 
not considered because the energy from the elastic collision was below the detector threshold. 

The inelastic processes considered in this analysis originate from spin-independent nuclear recoil events that produce either a photon or an electron \cite{Migdal_INSS, Brem_KP}.
Therefore, these results are directly comparable to existing limits for DM-nucleon interactions.

In this paper, we present a re-analysis of data from the Super Cryogenic Dark Matter Search (SuperCDMS) experiment to look for DM scattering inelastically off of nuclei.
\Cref{sec:SCDMS} describes the experiment and data selection.
\Cref{sec:Velocity_Damping} discusses how we account for the scattering of DM through the atmosphere and Earth before it reaches the underground experiment.
\Cref{sec:Signals} details the two signal models considered in this analysis.
\Cref{sec:BgModels} specifies the background models included in the likelihood framework, and \Cref{sec:Likelihood} describes the limit-setting method.
The final results are presented in \Cref{sec:Results}.


\section{SuperCDMS}\label{sec:SCDMS}

The SuperCDMS experiment was operated $\sim$~700 m underground in the Soudan Underground Laboratory from 2012 to 2015.
During this period, 15 germanium crystal detectors were used to search for DM particle masses from a few to tens of GeV$/c^2$~\cite{CDMS_HT,CDMSlite_long,CDMSlite_R3}. 
The 3 inch diameter, 1 inch thick cylindrical detectors were shielded from ambient radiation in the experimental cavern by layers of polyethylene, lead, and a copper cryostat.
The crystals were instrumented with interleaved Z-sensitive Ionization and Phonon (iZIP) sensors~\cite{iZIP_discrimination}. 
The detectors were biased face-to-face with $\pm~2~\textrm{V}$ and achieved an electron-recoil energy threshold of about 860 eV with discrimination between nuclear recoils (NR) and electron recoils (ER) 
down to $2~\textrm{keV}$~\cite{CDMS_LT}.

Two detectors were, for some periods, operated with a high-voltage bias near $\sim$75~V across the crystal~\cite{CDMSlite_long,CDMSlite_R3}.
This mode of operation, referred to as the CDMS low ionization threshold experiment (CDMSlite), takes advantage of the Neganov-Trofimov-Luke (NTL) mechanism~\cite{Neganov_Trofimov,Luke} to amplify small ionization signals. 
The amplification lowers the threshold of the experiment below an energy of  $100~\textrm{eV}_{\textrm{ee}}$ (ER equivalent energy)~\cite{CDMSlite_long,CDMSlite_R3}, but
sacrifices all discrimination between NR and ER events.

\subsection{CDMSlite Data}

For this analysis, we consider the data collected by one of the CDMSlite detectors that was operated from February 2015 to May 2015 and collected $60.9~\textrm{days}$ of raw livetime~\cite{CDMSlite_R3}. The exposure was divided into two segments, Period 1 (P1) and Period 2 (P2), due to changes in the operating conditions and parasitic resistance that affected the actual voltage across the crystal. 
These data were first analyzed in Ref.~\cite{CDMSlite_R3}. Thus, the search presented in this paper was conducted on an unblinded dataset.

\subsection{Event Selection}

Since the energy region of interest for this analysis largely overlaps with Ref.~\cite{CDMSlite_R3}, the same data quality selection criteria were used to remove problematic events such as those arising from low-frequency mechanical noise, electronic glitches, and poorly reconstructed pulse shapes.

The grounded copper housing surrounding the detector distorts the electric field near the edges of the crystal. Since the electric potential is reduced in these regions, the amplification is not uniform throughout the crystal. 
To minimize the number of events with reduced amplification, the same fiducial volume selection as defined in Ref.~\cite{CDMSlite_R3} was adopted, which rejects events with low NTL amplification. After applying the selection criteria, the remaining exposure is $36.9~\textrm{kg} \cdot \textrm{d}$ and the analysis threshold is $70~\textrm{eV}_\textrm{ee}$~\cite{CDMSlite_R3}.


\section{Damped Velocity Distribution}\label{sec:Velocity_Damping}

Calculation of the DM flux requires knowledge of the velocity distribution of the incoming DM. 
The standard halo model (SHM) was assumed, which is based on a Maxwell-Boltzmann velocity distribution with a characteristic velocity of $220~\textrm{km/s}$ \cite{10.1093/mnras/221.4.1023}.
Particles traveling at velocities greater than the escape velocity of the galaxy are not gravitationally bound, so the distribution is truncated at $544~\textrm{km/s}$ and re-normalized \cite{10.1111/j.1365-2966.2007.11964.x}.
The local DM density is assumed to be $0.3~\textrm{GeV}/(c^2\; \textrm{cm}^3$) \cite{LEWIN199687}.

The Earth is typically considered to be transparent to DM. However, for large coupling strengths between DM and nuclei, on the order of $\sim$$10^{-30}~\textrm{cm$^2$}$, the Earth and atmosphere can no longer be neglected. 
As DM particles travel through the Earth and atmosphere, they can scatter off of atoms and lose energy.
In the most extreme case, the DM can lose enough energy so that it can no longer create a signal above the detector threshold.
This Earth shielding limits the sensitivity of experiments since DM with stronger couplings is fully attenuated before reaching the detector~\cite{terrestrial_effects}.

This analysis accounts for the attenuation effect by damping the DM velocity distribution, described as ``Method B" in Ref.~\cite{blind_underground}. This approach allows for DM with cross sections in an intermediate region to scatter and lose some energy yet still reach the detector. It is also flexible enough to use a more complex shielding model, which is detailed in \Cref{sec:Shielding}.

Interactions with normal matter alter the velocity of DM particles at the detector by shifting the distribution to lower values. The DM velocity distribution at the detector site is calculated from the average energy loss via scattering off of nuclei, as described in Ref.~\cite{blind_underground,strong_DM}.
Since we concern ourselves with light dark matter in this paper, we assume a nuclear form factor of unity. 
A velocity damping parameter, $\kappa$, is defined as:
\begin{equation}
    \kappa \equiv \frac{\rho \sigma_n}{m_{\chi} \mu_n^2} \left( \sum_i^{\mathrm{elements}} \frac{F_i \mu_i^4 A_i^2}{m_i^2} \right) d \,,
\label{eq:kappa}
\end{equation}
with the variables defined in \Cref{tab:veloparam}~\cite{blind_underground}. 
The calculation of $\kappa$ is location specific since it depends on the path length, $d$, and the type of material the particle travels through before reaching the detector.

\begin{table}[htpb]
\centering
\caption{Definition of variables in \Cref{eq:kappa} to calculate the velocity damping parameter, $\kappa$.}
    \begin{tabular}{|c|l|}
    \hline
    Variable     & Definition\\
    \hline
    $\rho$          & material density \\
    $\sigma_n$      & DM-nucleon scattering cross section \\
    $\mu_n$         & reduced mass (DM particle, nucleon)\\
    $i$             & element index \\
    $F_i$           & mass fraction of the element $i$ \\
    $\mu_i$         & reduced mass (DM particle, nucleus of element $i$)\\
    $A_i$           & atomic number of element $i$\\
    $m_i$           & atomic mass of element $i$\\
    $d$             & path length through the shielding layer\\
    \hline
    \end{tabular}

\label{tab:veloparam}
\end{table}

An incoming DM particle with initial velocity $v_i$ reaches the detector with velocity $v_f$, 
\begin{equation}
	v_f = v_i e^{-\kappa} \,,
\label{eq:damp}
\end{equation}
thus the DM velocity distribution at the detector ($f_d$) is given by:
\begin{equation}
	f_d\left( v_f \right) = e^{2 \kappa} \times f\left(v_f e^{\kappa} \right) \,,
\label{eq:vtran}
\end{equation}
where the incoming SHM velocity distribution ($f)$ has been transformed by the effects of shielding~\cite{blind_underground}.
The exponential term in front of the distribution accounts for the normalization of the increased flux caused by the attenuated velocity distribution.

Implicit in the definition of $\kappa$ is the dependence on the incident angle of the DM with respect to the detector.
For example, particles originating from directly above the experiment pass through the atmosphere, the local overburden, and the experimental shielding. Meanwhile, a particle from below the experiment will traverse the internal structure of the Earth instead of the local overburden. Therefore, the incoming DM flux must be evaluated for every angle. More information about the angular dependence can be found in \Cref{sec:angular_dependence}.
We assume that the particles travel in straight line trajectories even though scattering will affect their trajectories. 
According to Ref.~\cite{blind_underground}, this approach still underestimates the number of dark matter particles with sufficient energy to interact with the detector.

\subsection{Earth Shielding}\label{sec:Shielding}

Due to the exponential nature of \Cref{eq:vtran}, $\kappa$ can be calculated for each layer of shielding independently using \Cref{eq:kappa} and summed to derive a cumulative value.
Four categories of shielding were considered for the SuperCDMS experiment at Soudan: the atmosphere, the overhead rock, the experimental shielding, and the bulk of the Earth below the experiment.

The density of the Earth's atmosphere decreases continuously as a function of altitude. However, for the purpose of simplifying the model, a seven layer atmosphere model based on the 1976 US Standard Atmosphere was used~\cite{atmosphere_model}. The value used for the density of each layer corresponds to the lowest altitude of that layer, thus overestimating the amount of shielding. Densities and altitudes from sea level are listed in \Cref{tab:atmos}.

\begin{table}[htbp]
   \begin{tabular}{|c|c|c|}
    \hline
    \rule{0pt}{10pt}Layer & Height range [km] & Maximum density [ g/m$^3$ ]  \\
   \hline
    1   &   0-11    &  1225    \\
    2   &   11-20   &  363.91  \\
    3   &   20-32   &  88.03   \\
    4   &   32-47   &  13.22   \\
    5   &   47-51   &  1.43    \\
    6   &   51-71   &  0.86    \\
    7   &   71-100  &  0.064    \\
    \hline
    \end{tabular}
\caption{Densities and maximum heights of the layers in the atmosphere model~\cite{atmosphere_model}.}
\label{tab:atmos}
\end{table}

The SuperCDMS experiment at Soudan was located under $714~\textrm{m}$ of Ely greenstone and iron ore in northern Minnesota. We consulted with a geologist at the University of Minnesota to obtain rock and chemical compositions of the Soudan region~\cite{UMNgeo}. Data were provided for sectors in eight geographic directions between radii of 100, 500, 1000, 5,000, 10,000, 20,000 and 50,000 meters.\footnote{
Geological data for the Soudan region is provided in two auxiliary files: the \texttt{SoudanRegion.csv} file contains the area and fraction of each rock type, and the elemental mass fractions for the chemical composition are described in \texttt{RockChem.csv}.
}
The composition and density of rock depends on the direction. 
To simplify the calculation and ensure that the amount of shielding is not underestimated, we select the direction that gives the maximum value of $\kappa$ for each mass and cross section considered.

The dominant component of the shielding is the Earth below the experiment.
The conventional model of the Earth is used, consisting of four concentric spheres: crust, mantle, outer core, and inner core. Details of the parameters defining the thickness and composition of each layer in the model are available in the appendix.

The smallest contribution to the damping model comes from the shielding around the experiment itself, which was approximated with concentric spheres. 
The outermost layer is $50~\textrm{cm}$ of polyethylene (C$_2$H$_4$) with a density of $0.94~\textrm{g/cm$^3$}$, followed by $22.5~\textrm{cm}$ of Pb and $3~\textrm{cm}$ of Cu~\cite{CDMS_II}.
The contribution to the velocity damping parameter from the shielding around the experiment is 0.2\% for a 1 $\textrm{GeV}/c^2$ DM particle with a nucleon scattering cross section of $10^{-30}~\textrm{cm}^2$ and traveling straight downward, where the Earth's shielding contribution is the weakest.

\subsection{Angular Dependence}\label{sec:angular_dependence}

The depth of the experiment in the Earth's crust produces an asymmetric angular distribution of shielding. Particles originating from below the experiment must traverse the majority of the Earth's diameter, while particles originating from directly above are only affected by the local overburden at the Soudan Mine.

This analysis follows the angular convention in Ref.~\cite{angular_coordinates}, which defines $\theta$ as the incident angle between the incoming DM particle's velocity and zenith at the experiment. By this definition, $\theta = 0$ corresponds to particles originating from directly below the experiment.
The total path length is calculated as:
\begin{equation}
    \ell = r_d \cos\theta + \sqrt{r_E^2 - \left( r_d \sin\theta \right)^2 } \,,
\label{eq:pathlength}
\end{equation}
where $r_d$ is the distance from the center of the Earth to the SuperCDMS experiment. The average value of $6471~\textrm{km}$ was used for the Earth's radius, $r_E$. The total path length can be generalized to determine $d$, the path length through individual layers, used in \Cref{eq:kappa}.

\begin{figure}[htpb]
    \includegraphics[width=\textwidth]{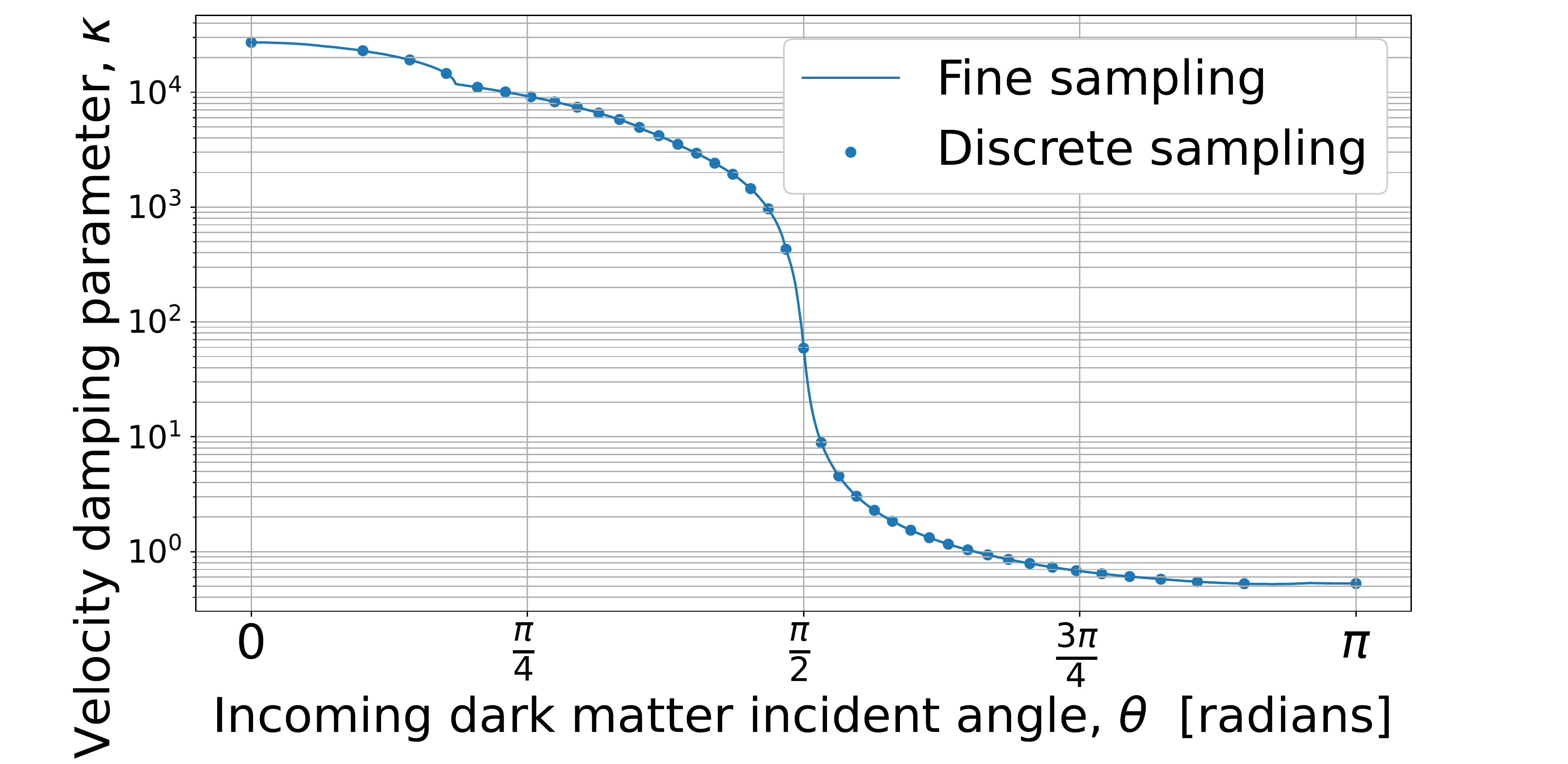}
    \caption{Value of $\kappa$ as a function of incident angle, $\theta$, for a DM particle with a mass of $1~\textrm{GeV}/c^2$ and a scattering cross section of $10^{-30}~\textrm{cm}^2$. The curve indicates the distribution and the dots indicate values at which the curve was sampled to calculate the signal models. The discontinuity below $\pi/2$ indicate the location of the transition between the core and the mantel while the more detailed overburden model leads to a smooth curve above $\pi/2$. }
    \label{fig:kappa_angle}
\end{figure}

Since the composition of the Soudan geology is well understood, the overburden above the experiment was modeled using the local geometry data and the shielding below the experiment was modeled according to the conventional Earth model. 
This introduced an angular dependence on the calculation of $\kappa$ that affects the velocity distribution at the detector. 
The signal model was integrated over the incoming angle, $\theta$, at twenty discrete points sampled uniformly in $\cos\theta$ as indicated in \Cref{fig:kappa_angle}.
We ignore the relationship between the WIMP wind and the Earth reference frame and calculate a single velocity distribution, which is attenuated according to its path through the shielding. 

\section{Inelastic Scattering Signals}\label{sec:Signals}

In this paper, we report on the search for a signal from DM interactions through two inelastic scattering channels. The first search channel is through bremsstrahlung radiation, where a photon is produced during the DM-nucleon scattering process.
The second search channel is induced by the Migdal effect, where a low-energy NR perturbs the atomic electron cloud, occasionally emitting electrons and/or photons.

\subsection{Bremsstrahlung Radiation}\label{sec:Brems}

The differential scattering rate for emitting a photon of energy $E_\gamma$ through the bremsstrahlung process has been derived by Kouvaris and Pradler in Ref.~\cite{Brem_KP}:
\begin{equation}
	\frac{d\sigma}{dE_\gamma} = 
	\frac{4 \alpha |f(E_\gamma)|^2}{3 \pi E_\gamma} \frac{\mu_N^2 v^2 \sigma_N}{m_N^2} \sqrt{1 - \frac{2 E_\gamma}{\mu_N v^2}} \left( 1 - \frac{E_\gamma}{\mu_N v^2} \right) \,,
\label{eq:brem1}
\end{equation}
where $\alpha$ is the fine structure constant, $f$ is the atomic scattering function discussed in \Cref{PACS}, $\mu_N$ is the reduced mass of the DM-nucleus system, $m_N$ is the mass of the nucleus, $v$ is the velocity of the incoming DM particle relative to the detector, and $\sigma_N$ is the interaction cross section between the DM particle and the nucleus. 
A signal spectrum is obtained by integrating over the velocity distribution at the detector while accounting for the angular dependence and the flux,
\begin{equation}
    \frac{dR}{dE_\gamma} = N_T \frac{\rho_{\chi}}{m_{\chi}} \int_{|\vec v|\geq v_{\rm min} } d^3\vec v\, v f_d(\vec{v}) \frac{d\sigma}{dE_\gamma} \,,
\label{eq:brem2}
\end{equation}
where $N_T$ is Avogadro's number divided by the atomic mass. 
The minimum velocity required to induce a recoil of $E_\gamma$, $v_{\rm min }$, is given by 
\begin{equation}
    v_{\rm min} = \sqrt{ \frac{m_N E_R}{2\mu^2_N} }  + \frac{E_\gamma}{ \sqrt{2 m_N E_R}},
    \label{eq:vmin}
\end{equation}
where $E_R$ is the nuclear recoil energy.

\subsubsection{Photoelectric Absorption Cross Section Uncertainty}\label{PACS}

The atomic scattering function, $f$, in \Cref{eq:brem1} is the sum of a real and complex portion,
\begin{equation}
    |f|^2 = |f_1 + i f_2|^2 = f_1^2 + f_2^2 \,.
\label{eq:bremF}
\end{equation} 
The components, $f_1$ and $f_2$, are defined in Ref.~\cite{xray_db} as:
\begin{equation}
    f_1 = Z^* + \frac{1}{ \pi r_e hc} \int_{0}^{\infty} \frac{\epsilon^2 \sigma_{a}(\epsilon) }{E_\gamma^2 - \epsilon^2}d\epsilon \,;
    \quad f_2 = \frac{\sigma_a}{2r_e \lambda} \,,
\label{eq:bremF1}
\end{equation}
for the nuclear contribution, 
where $r_e$ is the classical radius of the electron, $\sigma_{a}$ is the photoelectric absorption cross section,  $\lambda$ is the wavelength of the emitted photon and $\epsilon$ is a variable of integration. 
$Z^*$ is defined as $Z -(Z/82.5)^{2.37}$, where $Z$ is the proton number.

\begin{figure}[htpb]
    \includegraphics[width=\textwidth]{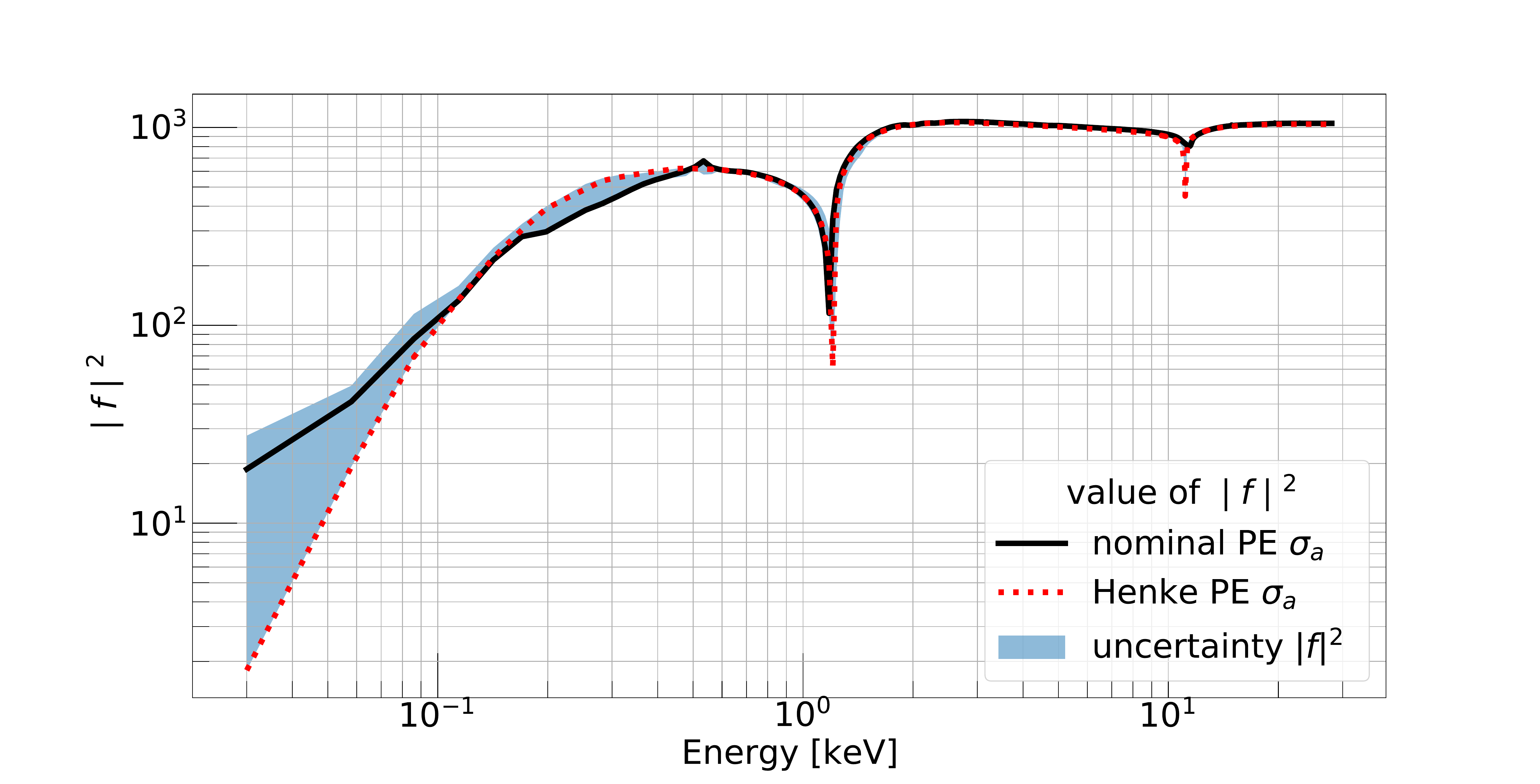}
    \caption{The atomic scattering function, $|f|^2$, calculated as a function of energy using different values of the photoelectric absorption cross section, $\sigma_a$, in germanium. The nominal value of $\sigma_a$ obtained from literature was used to calculate the solid black curve. 
    The range of uncertainty on $|f|^2$ is indicated by the shaded region.
    }
    \label{fig:ASF}
\end{figure}

Measurements of $\sigma_a$ at low energies have a range of values which lead to significant systematic uncertainties of its value~\cite{CDMS_HVeV}.
Both $f_1$ and $f_2$ depend on $\sigma_a$, so this uncertainty enters into the calculation of the expected event rate.
Based on existing literature~\cite{PE_Henke,PE_2,PE_5,PE_9,PE_12,PE_24,PE_26,PE_36,PE_38,PE_39,PE_42,PE_43,PE_44,PE_47,PE_NIST}, 
a nominal value and uncertainty band for $\sigma_a$ were derived as a function of energy.
Using the resulting values of $\sigma_a$, $|f|^2$ was calculated and compared in \Cref{fig:ASF}, along with the commonly referenced Henke dataset~\cite{PE_Henke}.
The range of variation is on the order of 30\% and is most prominent at low energies. 
The signal calculation uses the lower bound from all calculations of $|f|^2$, 
that conservatively predicts a lower expected signal rate and thus results in a weaker limit.

\subsection{Migdal Effect}\label{sec:Migdal}

When a NR occurs, there is a delay between the initial recoil and the response of the surrounding electron cloud, effectively boosting the entire electron cloud simultaneously
with respect to the nucleus
~\cite{Migdal_Approx}.
The displacement of the nucleus due to a DM scattering event dramatically changes the wavefunction of the electrons in the surrounding electron cloud. As the electron cloud relaxes back to the ground state, an electron can transition to a free state (i.e.~be ejected), a process known as the Migdal effect~\cite{Migdal_Approx}. 
The formulation of this effect as applied to DM direct detection has been calculated by Ibe, Nakano, Shoji, and Suzuki~\cite{Migdal_INSS}. 
An alternative approach utilizing the photoelectric cross section has been developed by 
Liu, Wu, Chi, and Chen \cite{Liu:2020pat}.
The Ibe {\it et al.} formulation was then applied to data by Dolan, Kahlhoefer, and McCabe~\cite{Migdal_DKM}.
In this paper, we adapt the Ibe {\it et al.} formalism to be consistent with other results within the community.

The differential rate for this process can be expressed as:
\begin{equation}
	\frac{\mathrm{d}^3 R_\text{ion}}{\mathrm{d}E_\mathrm{R}\,\mathrm{d}E_e\,\mathrm{d}v} =
	\frac{\mathrm{d}^2R_\text{nr}}{\mathrm{d}E_\mathrm{R}\,\mathrm{d}v} \times 
	\sum_{nl} \frac{1}{2 \pi} 
	\frac{\mathrm{d} p^c_{q_e}(nl \to E_e)}{\mathrm{d}E_e} \,,
\label{eq:mig1}
\end{equation}
where $p_{q_e}^c(nl \to E_e)$ is the transition probability for an electron, with momentum $q_e = m_e v_{\mathrm{ nucleus}}$ with respect to a stationary nucleus, to be ejected from quantum state $nl$ with energy $E_e$, and $\mathrm{d}^2 R_\text{nr}/(\mathrm{d}E_\mathrm{R}\,\mathrm{d}v)$ describes the incoming DM flux and scattering rate,
\begin{equation}
	\frac{\mathrm{d}^2 R_\text{nr}}{\mathrm{d}E_\mathrm{R}\,\mathrm{d}v} = \frac{\rho_\chi \, \sigma_N}{2 \, \mu_N^2 \, m_{\chi} } \frac{f_d(v)}{v} \,.
\label{eq:mig2}
\end{equation}
The factor of 1/2${\pi}$ in \Cref{eq:mig1} is a normalization constant and is consistent with the formulation of \cite{Migdal_INSS}.
The transition probability tables that were provided by Ref.~\cite{Migdal_INSS} have been evaluated at a reference velocity, $v_{\mathrm{ref}} = 10^{-3}c$. 
Conversion between the reference value and an arbitrary electron momentum is given by:
\begin{equation}
    p_{q_e}^c \left( nl \rightarrow E_e \right) = \left( \frac{q_e }{ v_{\mathrm{ref}} m_e } \right)^2 p_{v_{\mathrm{ref}}}^c \left( nl \rightarrow E_e \right) \,.
    \label{eq:ibe1}
\end{equation}

These tables were calculated assuming an isolated atom.
In a crystal detector, atoms are not isolated and the electron clouds are subject to interactions with atoms in nearby lattice sites, in particular, the outer electron shells.
We assumed crystal effects are negligible for the inner shells and exclude the outermost germanium shell ($n=4$) from the analysis \cite{PhysRevLett.127.081804,PhysRevLett.127.081805,Liang:2019nnx}.
The inner electron shells are assumed to be dominated by interactions with the nucleus and are sufficiently representative of an isolated atom.
Excluding the valence shell causes a minimal decrease in the expected signal rate because most of the signal originates from the inner shells; the majority of the valence shell contribution is below the experimental threshold.
Excluding part of the signal model results in a more conservative estimate of the expected rate.

\begin{figure}[htpb]
  \centering
  \includegraphics[width=\textwidth]{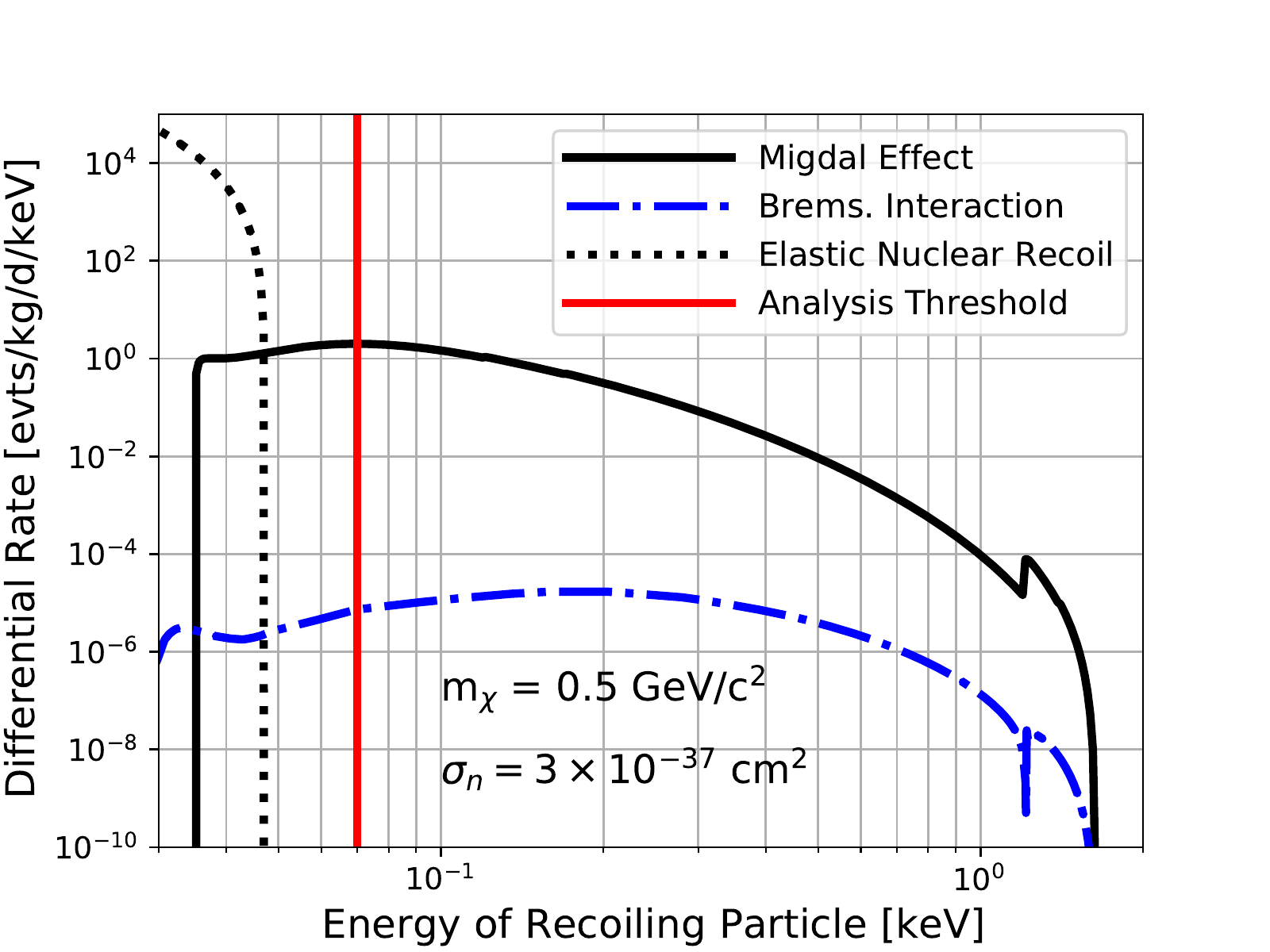}
  \caption{Comparison of the expected DM signal rates in germanium. For the DM mass chosen, $0.5~\textrm{GeV}/c^2$, the elastic NR signal is  below the analysis threshold of $70~\textrm{eV}$. The Migdal signal extends to higher energies, but at a smaller rate than the NR signal. The Migdal signal has a sharp cutoff at low energies because the valence shell was not included in the analysis. The bremsstrahlung signal extends to the same energy as the Migdal, but with a smaller expected rate.  The bump above $1~\textrm{keV}$ is where the $n=2$ electron shell starts contributing to the signal rate. }
  \label{fig:signals}
\end{figure}

\Cref{fig:signals} shows the expected signal rates for an incident DM particle with $0.5~\textrm{GeV}/c^2$ mass and $10^{-35}~\textrm{cm}^2$ cross section. This mass was chosen to highlight the sub-GeV reach of the inelastic channels where the elastic NR signal is below the energy threshold of the analysis and thus undetectable. 
The chosen cross section is small enough that the Earth's shielding is a negligible effect.

\section{Background Modeling and Systematics}\label{sec:BgModels}

In order to perform a profile likelihood analysis, all background sources must be well understood and modeled. The background sources considered in this analysis are: neutron activation by the \textsuperscript{252}Cf calibration source, cosmogenic activation of the crystal, radiogenic Compton scattering, and \textsuperscript{210}Pb surface contamination. Inelastic scattering of low-energy neutrons was also considered as a background source, but it was determined via simulation that the background contribution was $\ll 1$ event for the exposure, and therefore negligible.

\subsection{Bulk Background Models}\label{sec:bulk_bg}

The dominant background originates from activation of the germanium crystal via the neutron calibration source. When stable \textsuperscript{70}Ge in the crystal captures a neutron, it becomes unstable \textsuperscript{71}Ge that decays via electron capture to \textsuperscript{71}Ga. The electron capture creates a cascade of energy in the form of  X-rays and Auger electrons, where the energy emitted from the decay is equal to the binding energy of the shell that captured the electron. The K-shell has the highest probability of capture at $87.57\%$ and results in a $10.37~\textrm{keV}$ emission. Each successive shell has a lower probability and emits less energy: the L-shell captures an electron $10.53\%$ of the time and emits $1.3~\textrm{keV}$, and the M-shell captures an electron $1.78\%$ of the time and emits $160~\textrm{eV}$~\cite{Pepin_thesis}. This background was modeled by a Gaussian distribution centered on each electron shell peak energy. The amplitudes of these peaks were set relative to the K-shell peak as determined by the capture probabilities of each shell. There is one overall normalization parameter  for the \textsuperscript{71}Ge background. 
Similarly, \textsuperscript{68}Ge decays to \textsuperscript{68}Ga which can decay through electron capture or through beta decay. 
The \textsuperscript{68}Ga beta endpoints are substantially higher than the energy range considered in this analysis so its contribution is neglected since we expect just 0.001 events.

Before the detectors are brought underground, cosmic-ray spallation can knock nucleons out of the germanium atoms in the crystal and create radioisotopes. One of the more problematic cosmogenic isotopes is tritium (\textsuperscript{3}H), which provides a persistent source of betas due to its half-life of $12.32~\textrm{years}$. The tritium background was modeled by a standard beta emission spectrum with an endpoint energy of $18.6~\textrm{keV}$ and an unconstrained normalization.
The resulting normalization from the fit was 50 $\pm$ 20 events, which is consistent with a previous dedicated analysis~\cite{Fascione_thesis,CDMSlite_tritium}.

Tritium decays dominate cosmogenic background rates, but spallation can also leave other unstable nuclei. The other residual nuclei were considered background sources if they have a half-life that is long enough that they will have not decayed away before data taking began, but also short enough that the activity is comparable to other background rates. 
The additional isotopes modeled in this analysis were \textsuperscript{68}Ga,
\textsuperscript{65}Zn, and \textsuperscript{55}Fe~\cite{CDMSlite_tritium}. Other isotopes considered are \textsuperscript{57}Co, \textsuperscript{54}Mn, and \textsuperscript{49}Vn, but the expected contribution of each was determined to be $<1$ event for the given exposure, and are neglected. 
Each of the modeled isotopes decays via electron capture, like the activated germanium, but at different energies.  
Contributions from K, L, and M-shells were modeled by Gaussian distributions with fixed relative amplitudes with a single normalization parameter, analogous to how \textsuperscript{71}Ge was treated above.

All of the radioisotopes, created cosmogenically or by source activation, described in this subsection are distributed nearly uniformly
throughout the detector volume\footnote{Studies based on simulation have shown slightly more bias towards the surface of the crystals due to self-shielding effects.}.
Therefore, the efficiency of the physics selection criteria that were developed for a uniform DM signal could also be applied to the modeling of the backgrounds originating from those isotopes.

Gamma rays emitted by long-lived naturally occurring unstable radioisotopes
typically have energies much greater than those of interest for this analysis, but high-energy photons can undergo Compton scattering.
To first order, this creates a flat background continuum throughout the analysis energy region,  
although the model also includes low-energy ``steps" at the electron binding energies. 
These steps occur when the energy deposited by a scattering photon has enough energy to overcome the binding energy of a particular electron shell.
As the amount of energy increases, the number of available electrons to scatter against increases and so does the corresponding interaction rate. 
This Compton background spectrum was modeled as a flat contribution with an error function at each shell energy, with a width corresponding to the detector resolution,  to model the steps~\cite{Barker_thesis,CDMSlite_R3}.
The relative step amplitudes were determined from an independent fit to simulation data, and the overall normalization was allowed to float in the likelihood function~\cite{CDMSlite_R3}.

\subsection{Surface Background Models}\label{sec:surf_bg}

Radon daughters plating out on the detectors or surrounding copper housing were treated differently than the bulk contamination described in \Cref{sec:bulk_bg}.
Although decays can implant radon daughters below the surface, they are predominantly classified as surface events. 
To model the expected experimental signature of these surface events, a Geant4 \cite{AGOSTINELLI2003250,ALLISON2016186,1610988} simulation of \textsuperscript{210}Pb surface contamination on  a tower of six germanium detectors was performed.
 The simulation allowed for the subsequent alpha decays to implant the long-lived \textsuperscript{210}Pb and mimic the physical radiocontamination~\cite{Barker_thesis}.
The simulation indicated that surface events in the detectors originated from three locations with direct line-of-sight: the top lid (TL) of the copper housing that directly faces the top detector in the tower, the sidewall housing (SH) around the outer radial wall of each detector, and the surfaces of the germanium crystal (GC) from both the detector itself and the face of the adjacent detector.

The simulated spectra were normalized using an independent measurement of the rate of alpha events in each of the detectors~\cite{Barker_thesis}. For the energy range used in this analysis, the normalization to the number of expected events from each contribution were determined to be
$N_{\mathrm{TL}} = 158.4 \pm 16.6$, $N_{\mathrm{SH}} = 22.3 \pm 1.4$, and $N_{\mathrm{GC}} = 23.5 \pm 5.9$ events, which contribute to the total number of background events in the likelihood. 
The housing copper is not as radiopure as the crystals so it dominates the contribution. 
Events originating from the bottom tower lid are shielded by the other detectors in the housing stack. 
Correlations between normalization and shape uncertainties are taken into account using morphing parameters as described in Ref.~\cite{CDMSlite_R3}.

\subsection{Efficiency Model}

All of the background and signal spectra were convolved with the overall efficiency of the data selection criteria; the details of which can be found in Ref.~\cite{CDMSlite_R3}.
The trigger efficiency is 90\% at $70~\textrm{eV}_\textrm{ee}$ and 100\% above $90~\textrm{eV}_\textrm{ee}$.
The data quality selection criteria efficiency approaches 100\% around $150~\textrm{eV}_\mathrm{ee}$.
The largest loss of efficiency at low energies is due to the fiducial volume selection, which passes roughly 60\% of the events above $200~\textrm{eV}_\mathrm{ee}$ and has almost zero efficiency at $70~\textrm{eV}_\mathrm{ee}$.

\begin{figure}[htpb]
  \centering
  \includegraphics[width=\textwidth]{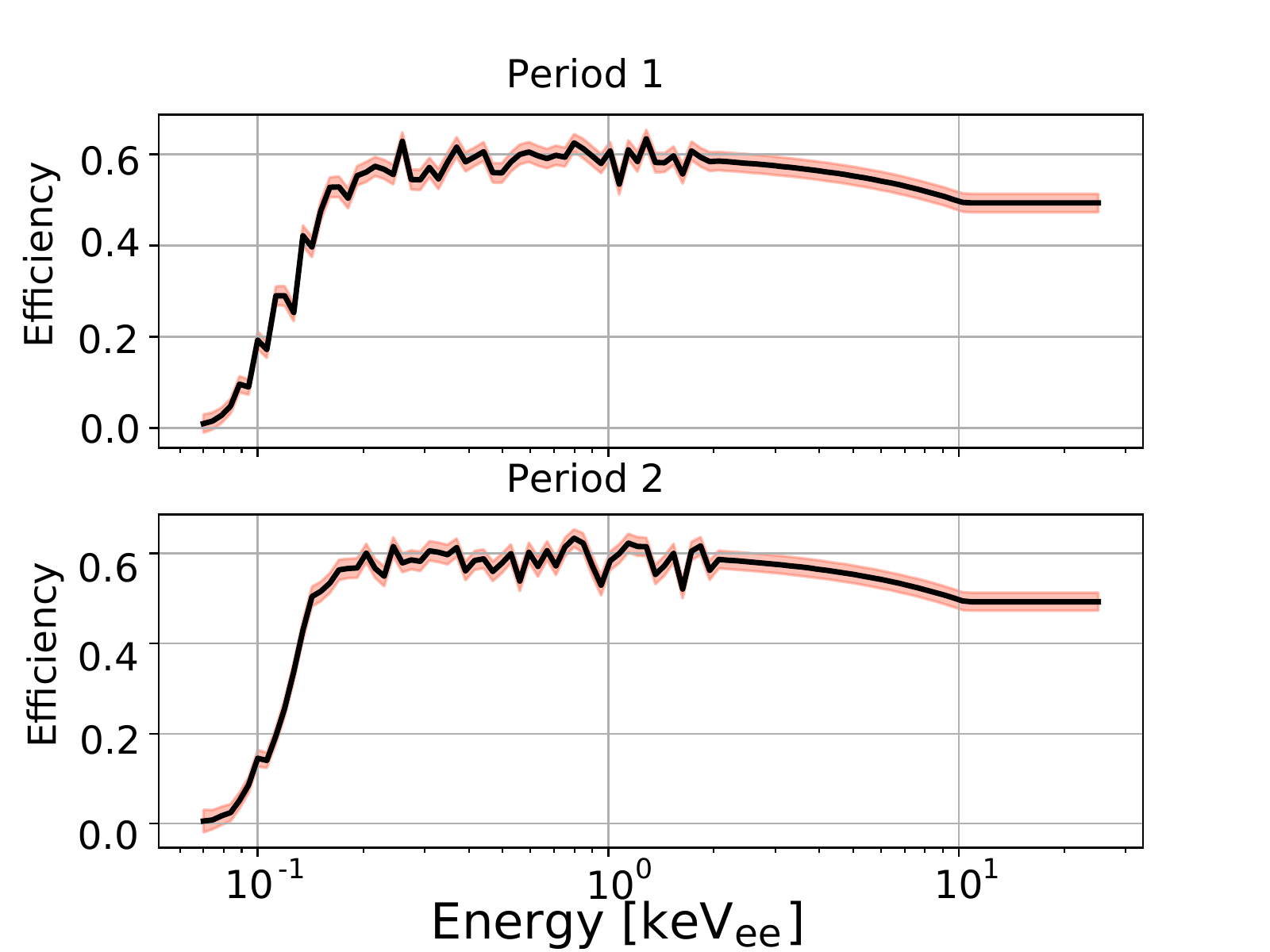}
  \caption{Efficiency of the data selection criteria in Ref.~\cite{CDMSlite_R3} for both CDMSlite data periods. The black curves are the median efficiencies and the red band indicates the $1\sigma$ uncertainty band.}
  \label{fig:Eff}
\end{figure}

\Cref{fig:Eff}\hphantom{ } shows the efficiency curve for each period used in this analysis. The notable fluctuations in the efficiency curve below $2~\textrm{keV}_\mathrm{ee}$ arise from the fiducial volume selection, which was calculated using simulated data with limited statistics. 
We take the efficiency curve with its statistical fluctuations. 
The efficiency models have been extended to $25~\textrm{keV}_\textrm{ee}$ following the procedure in Ref.~\cite{CDMSlite_tritium}.

The uncertainty on the efficiency curve was incorporated into the likelihood function via implementation of nuisance parameters in the maximum likelihood fit, one for each data period. The nuisance parameter is not a simple normalization factor, but a morphing parameter that allows for correlated variation between shape and normalization of the efficiency function.
This is accomplished by constructing 1-sigma bands around the median parameterized as a Gaussian distribution
following the same prescription as the surface background models in Ref.~\cite{CDMSlite_R3}.

\subsection{Resolution Model}

The background model and signal models were convolved with the energy resolution of the detector, thus a resolution model was included in the likelihood function as additional nuisance parameters.
The functional form of the detector energy resolution is:
\begin{equation}
\label{eq:resolution_function}
    \sigma^2(E) = \sigma_E^2 + B E + (A E)^2 \,,
\end{equation}
where $\sigma$ is the resolution, $E$ is the total phonon energy, $\sigma_E$ is the baseline noise resolution that originates from the readout electronics, $B$ is a variance that scales with energy, and $A$ accounts
for any effects that scale with energy such as pulse shape variation and position dependence~\cite{CDMSlite_long}. 
The parameters used are described in Section II.D of Ref.~\cite{CDMSlite_R3}.
At 100 eV$_{\mathrm{ee}}$, the resolution of period 1 and 2 is 14 eV$_{\mathrm{ee}}$ and 16 eV$_{\mathrm{ee}}$ respectively. 
Near the upper end of the analysis range at 25 keV$_{\mathrm{ee}}$, the resolution is 193 eV$_{\mathrm{ee}}$ and 198 eV$_{\mathrm{ee}}$ for periods 1 and 2, respectively.

\subsection{Nuclear Recoil Ionization Yield}
The energy spectrum measured in the detector is a combination of the ER and NR components. 
In calculating the expected signal rate, an assumption about the nuclear recoil ionization yield, $Y(E_\mathrm{NR})$, is needed.
We adopt the Lindhard model (\Cref{eq:lindhard}),
\begin{equation}
 Y_{\text{Lindhard}}(E_{\mathrm{NR}}) = k\cdot g(\epsilon) /( 1 +  k\cdot g(\epsilon) ),     
\label{eq:lindhard}    
\end{equation}
where $g(\epsilon)= 3\epsilon^{0.15} + 0.7\epsilon^{0.6} +\epsilon$,  
$\epsilon = 11.5E_{\mathrm{NR}}({\mathrm{keV}})Z^{-7.3}$, 
$Z$ is the atomic number, and
$k$ is the free electron energy loss
\cite{CDMSlite_long,Lindhard:1964}.

There is evidence of deviations from the Lindhard model at low energies~\cite{Chavarria:2016xsi,PhysRevD.103.122003}.
Therefore, the ionization production was cut off at 22.7~eV, according to an independent measurement of the defect energy creation threshold~\cite{SuperCDMS:2018fip}. 
This is implemented with a hyperbolic tangent function, 
\begin{equation}
\label{eq:Yed_function}
\begin{aligned}
    Y(E_{\mathrm{NR}}) =&\frac{1}{2}Y_{\text{Lindhard}}(E_{\mathrm{NR}})\\
    &\times     \left(\tanh\left(\frac{E_{\mathrm{NR}}-22.7\; \mathrm{eV} }{22.32\; \mathrm{eV}}\right)+1\right),
    \end{aligned}
\end{equation}
that has a width of 22.32~eV, which was determined by requiring that the yield nearly vanish ($Y$=0.001) near the band gap energy of 0.74 eV.

Systematic uncertainties on the Lindhard model are propagated through its uncertainties in the {\it k} parameter.
For germanium, a nominal value of $k=0.157$ is assumed, with a Gaussian uncertainty of $\sigma = 0.05$~\cite{CDMSlite_R3}.
The signal model is calculated for the nominal, $\pm 1\sigma$, and $\pm 3\sigma$ values of {\it k} and intermediate values of $\sigma$ are interpolated.

\subsection{Analysis Energy Range}

\begin{figure}[htpb]
  \centering
  \includegraphics[width=\textwidth]{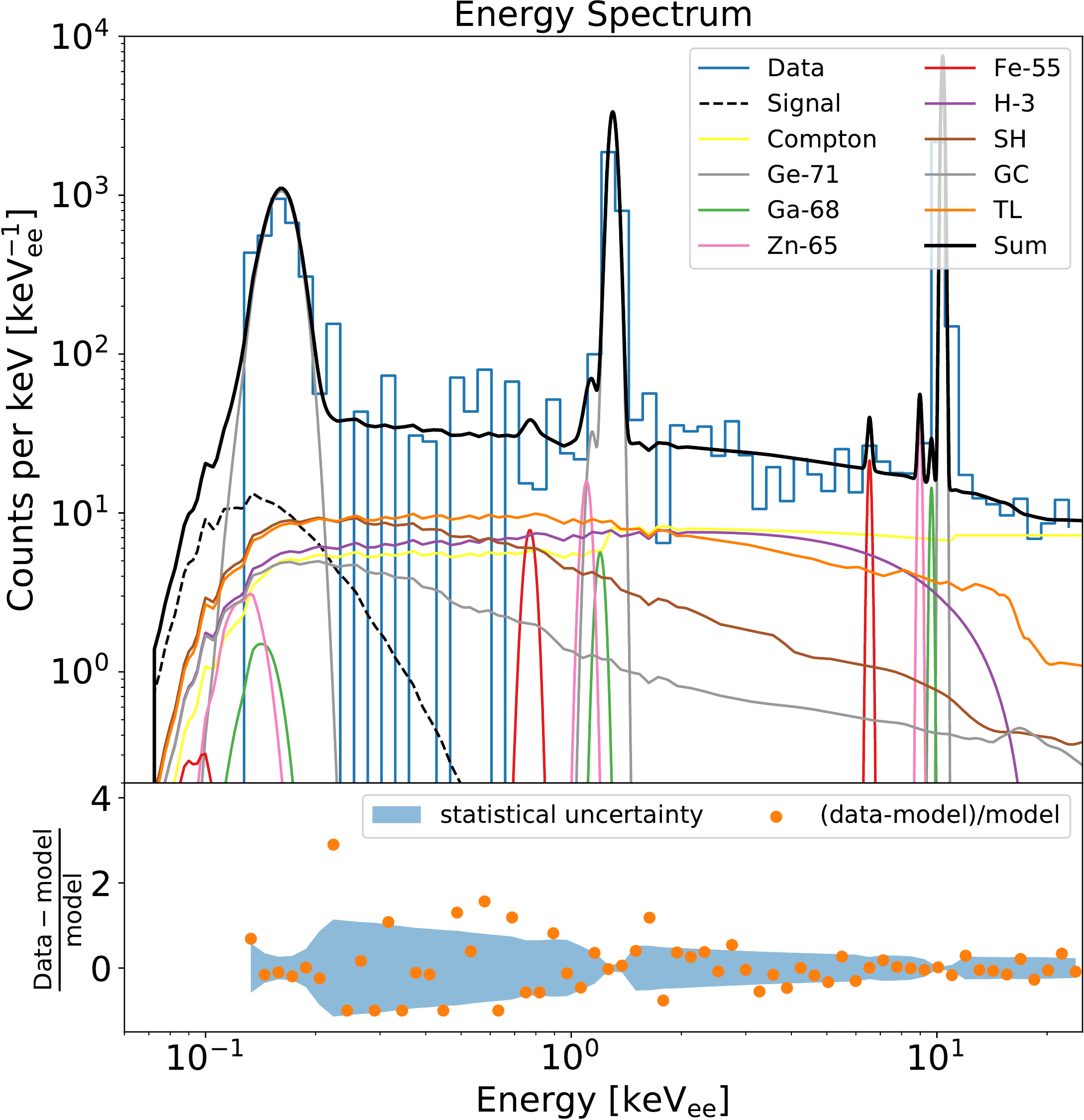}
  \caption{Example of an energy spectrum from the maximum likelihood fit for a Migdal signal model for a WIMP with a mass of 0.5 GeV/c$^2$ and a cross section of $3\times 10^{-37}$ cm$^2$ (black dashed curve). The data (blue histogram) have been logarithmically binned and overlaid with the background models (colored solid curves). The thick black line is the sum of all the models. 
  Normalization of the surface background model components (TL, SG and GC) are described in \Cref{sec:surf_bg}. 
  The plot on the bottom shows the residual between data and the model with the 1 $\sigma$ statistical uncertainty indicated by the shaded region. 
  }
  \label{fig:spectrum}
\end{figure}

The normalization of the models was determined by the fit to data, as described in~\Cref{sec:Likelihood}. 
An example of fitting the data is shown in~\Cref{fig:spectrum}, 
where the background models and Migdal signal model, for a WIMP with a mass of 0.5 GeV/c$^2$ and a cross section of 3$\times 10^{-37} \mathrm{cm}^2$,
have been scaled by the efficiency and convolved with the resolution model.

The shape of the background models dictated the energy range used in this analysis. A strong degeneracy between unconstrained models that are flat or nearly flat resulted in lack of fit convergence when maximizing the likelihood.
This degeneracy is broken by extending the analysis region to $25~\textrm{keV}_\textrm{ee}$ past the energy region considered in Ref. \cite{CDMSlite_R3}.
In contrast, the surface background rates were constrained by the independent measurement of alpha rates through Gaussian constraints in the likelihood function, and the cosmogenic activation lines were modeled with Gaussian distributions that are not degenerate with flat background models.


\section{Profile Likelihood Analysis}\label{sec:Likelihood}

An unbinned profile likelihood function was utilized for this analysis because it provides the ability to quantify an excess of events above the expected background and potentially claim a DM discovery.
In the absence of an excess, a likelihood places a more constraining exclusion limit than the Optimum Interval~\cite{Yellin_OI,Yellin_highstats} technique because it incorporates knowledge of the background. It also provides a rigorous and convenient way to account for systematic uncertainties in signal and background model parameters.

\subsection{Likelihood Function}

\Cref{tab:Lhood_variables} contains a list of the variables used in the likelihood function.
The likelihood function is composed of three types of terms.
The first type is the overall normalization 
term, ($\mathcal{L}_\textrm{Poiss}$), which allows the total number of fitted events to fluctuate around the number of events in the dataset ($N$). 
The second type ($\mathcal{L}_{\chi,bb,sb}^{P1}$, $\mathcal{L}_{\chi,bb,sb}^{P2}$) is the core of the likelihood that uses the signal and background PDFs to determine the most likely signal and background rates.
The third type constrains the nuisance parameters using auxiliary measurements. In this analysis, there are six terms: the morphed surface backgrounds ($\mathcal{L}_\textrm{surf}$), the morphed efficiency ($\mathcal{L}_\textrm{eff}^{P1}$, $\mathcal{L}_\textrm{eff}^{P2}$), the resolution model ($\mathcal{L}_\textrm{res}^{P1}$, $\mathcal{L}_\textrm{res}^{P2}$), and the yield model ($\mathcal{L}_\textrm{yield}$). 
The constraint terms are either a univariate Gaussian PDF, in the case of the efficiency and yield, or multivariate Gaussian distributions for the surface backgrounds and resolution. In order to implement a multivariate constraint, a covariance matrix is calculated from the normalization uncertainty of the individual components and the correlations between them~\cite{Cowan_book}.

\begin{table}[!htb]
  \caption{\label{tab:Lhood_variables}Definition of variables used in the likelihood function (\Cref{eq:Lhood_final}), sorted by their given state in the fit. Identifiers are used to label variables that are specific to a data period, iterators are used to distinguish items from a given set, free/constrained indicates the state in the fitting procedure, constants are inputs to the likelihood that were calculated in advance, and the signal and background models are functions of energy.}
  \begin{tabular}{|c|l|c|}
  \hline
  Variable            & Definition                                  &  State  \\
  \hline
  $P1$                & data period 1                               &  Identifier  \\
  $P2$                & data period 2                               &  Identifier  \\
  \hline
  $n$                 & iterator over $N$ events                    &  Iterator  \\
  $bb$                & bulk background iterator                    &  Iterator  \\
  $sb$                & surface background iterator                 &  Iterator  \\
  $i,j$               & general nuisance iterators                  &  Iterator  \\
  \hline
  $\nu_{\chi}$        & number of signal events                     &  Free  \\
  $\nu_{bb}$          & number of events in $bb$                    &  Free  \\
  $\nu_{sb}$          & number of events in $sb$                    &  Constrained  \\
  $s$                 & surface bg morphing parameter               &  Constrained  \\
  $\Xi$                 & efficiency morphing parameter               &  Constrained  \\
  $r$                 & resolution nuisance parameter               &  Constrained  \\
  $\theta_{Y}$        & yield nuisance parameter                   &  Constrained  \\
  \hline
  $N$                 & number of events in data                    &  Constant  \\
  $E_n$               & energy of event $n$                         &  Constant  \\
  $\mu_s$             & expected value of $s$                       &  Constant  \\
  $\mu_\Xi$             & expected value of $\Xi$                       &  Constant  \\
  $\mu_r$             & expected value of $r$                       &  Constant  \\
  $\sigma_\Xi$          & uncertainty of $\Xi$                          &  Constant  \\
  $\boldsymbol{S}$    & surface bg covariance matrix                &  Constant  \\
  $\boldsymbol{R}$    & resolution covariance matrix                &  Constant  \\
  \hline
  $f_{\chi}$          & signal PDF                                  &  Function  \\
  $f_{bb}$            & PDF of $bb$                                 &  Function  \\
  $\rho_{sb}$         & event density function of $sb$                       &  Function  \\
  \hline
  \end{tabular}
\end{table}

This analysis used the MINUIT algorithm~\cite{MINUIT} via the iminuit~\cite{iminuit} Python interface to maximize the log-likelihood function when evaluating the test statistic, which will be defined in \Cref{eq:Lhood_teststat_new}.

The full likelihood function used in this analysis is given by \Cref{eq:Lhood_final}, which is a single likelihood encompassing both data periods. 

\begin{figure}
\begin{equation}
\begin{gathered}
    \label{eq:Lhood_final}
    \ln\mathcal{L} = \\[10pt]
    \ln(\mathcal{L}_\textrm{Poiss})
    + \ln(\mathcal{L}_{\chi,bb,sb}^{P1})
    + \ln(\mathcal{L}_{\chi,bb,sb}^{P2})
    + \ln(\mathcal{L}_\textrm{surf}) \\
    + \ln(\mathcal{L}_\textrm{eff}^{P1})
    + \ln(\mathcal{L}_\textrm{eff}^{P2})
    + \ln(\mathcal{L}_\textrm{res}^{P1})
    + \ln(\mathcal{L}_\textrm{res}^{P2})
    + \ln(\mathcal{L}_\textrm{yield}) = \\[10pt]
  	- \left( \nu_{\chi} + \sum\limits_{bb=1}^6 \nu_{bb} + \sum\limits_{sb=1}^3 \nu_{sb} \right) \\
  	+ \sum_{n=1}^{N^{P1}} \ln\left[ \nu_{\chi} f_{\chi}^{P1}(E_n) + \sum\limits_{bb=1}^6 \nu_{bb} f_{bb}^{P1}(E_n) + \sum\limits_{sb=1}^3 \rho_{sb}^{P1}(E_n) \right] \\
  	+ \sum_{n=1}^{N^{P2}} \ln\left[ \nu_{\chi} f_{\chi}^{P2}(E_n) + \sum\limits_{bb=1}^6 \nu_{bb} f_{bb}^{P2}(E_n) + \sum\limits_{sb=1}^3 \rho_{sb}^{P2}(E_n) \right] \\
    - \frac{1}{2} \sum\limits_{i,j=1}^3 \left[ (s_i - \mu_{s_i})^T (\boldsymbol{S}_{i,j})^{-1} (s_j - \mu_{s_j}) \right] \\
    - \frac{{(\Xi^{P1} - \mu_{\Xi}^{P1})}^2}{2{(\sigma_{\Xi}^{P1})}^2}
    - \frac{{(\Xi^{P2} - \mu_{\Xi}^{P2})}^2}{2{(\sigma_{\Xi}^{P2})}^2} \\
    - \frac{1}{2} \sum\limits_{i,j=1}^{3} \left[ (r_i^{P1} - \mu_{r_i}^{P1})^T (\boldsymbol{R}_{i,j}^{P1})^{-1} (r_j^{P1} - \mu_{r_j}^{P1}) \right] \\
    - \frac{1}{2} \sum\limits_{i,j=1}^{3} \left[ (r_i^{P2} - \mu_{r_i}^{P2})^T (\boldsymbol{R}_{i,j}^{P2})^{-1} (r_j^{P2} - \mu_{r_j}^{P2}) \right] \\
    - \frac{1}{2} \theta_{Y}^2
    \,.
\end{gathered}
\end{equation}
\end{figure}

\subsection{Limit Calculation}

In a typical DM search, the number of signal events is directly proportional to the interaction cross section. Therefore, a constraint on the normalization of the signal model can be directly converted to a single cross section because there is a one-to-one mapping between cross section and number of signal events. 
In these searches, the shape of the signal model is determined by the mass of the DM particle since the velocity distribution is considered unchanged from space compared to the location of the experiment on Earth.

This assumption does not apply for any analysis that involves DM models with potentially large cross sections due to the shielding discussed in \Cref{sec:Velocity_Damping}. 
According to \Cref{eq:vtran}, the effects of the shielding shift the velocity distribution of DM particles at the detector site to lower values.
The moderated velocity distribution affects the expected DM-nucleon scattering rate and the shape of the recoil spectrum. 
This results in the shielding parameter, defined by \Cref{eq:kappa} and the shape of the expected signal to depend on the DM mass and cross section.

In order to test DM hypotheses in the mass and cross section parameter space, a test statistic based on the profile likelihood ratio was defined as:
\begin{equation}
    q(\nu_{\chi}) = \left\{ \begin{array}{cr}
    -2 \ln \left( \frac{\mathcal{L}(\nu_{\chi},\hat{\hat\theta},m_{\chi},\sigma_n)}{\mathcal{L}(\hat{\nu}_{\chi},\hat{\theta},m_{\chi},\sigma_n)} \right)  &  \nu_{\chi} > \hat{\nu}_{\chi} \\
    0  &  \nu_{\chi} < \hat{\nu}_{\chi}
    \end{array} \right. \,,
\label{eq:Lhood_teststat_new}
\end{equation}
where $\nu_{\chi}$ is the number of signal events, and $\theta$ is the vector of nuisance parameters.
The number of signal events given by the global likelihood maximum ($\hat{\nu}_{\chi}$) corresponds to the best fit values of the nuisance parameters ($\hat\theta$). 
The best fit values of the nuisance parameters when the signal model is fixed is indicated by $\hat{\hat{\theta}}$.
In order to calculate an upper limit, $q$ was set to zero when the number of signal events being tested was lower than the best fit number of signal events.

\begin{figure}[htpb]
  \centering
  \includegraphics[width=\textwidth]{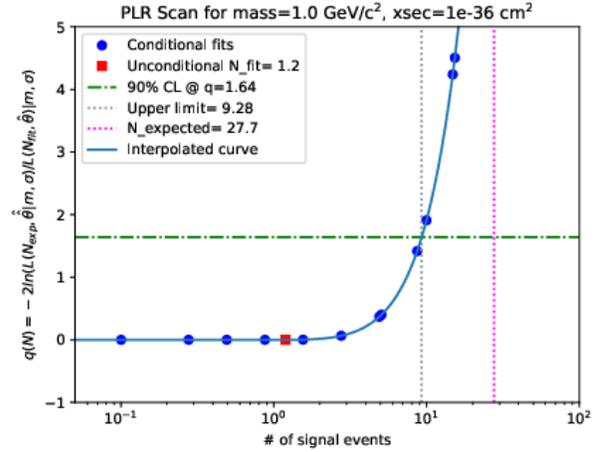}
  \caption{Example of a profile likelihood ratio (PLR) scan of $q$ for a given mass and cross-section point, with the number of expected signal events given by N\textunderscore expected.
  The upper limit on the number of signal events is determined by the position where the ratio crosses 1.64. 
  }
  \label{fig:PLRscan}
\end{figure}

\Cref{eq:Lhood_teststat_new} was evaluated over a grid of mass and cross section points.
At each point being tested, the signal shape was held constant and the number of signal events ($\nu_{\chi}$) was increased until the value of $q$ exceeded 1.64 (as shown by \Cref{fig:PLRscan}), which corresponds to the 90\% confidence level based on Wilk's theorem. 
If the upper limit was greater than the predicted number of signal events, then the signal hypothesis is consistent with the data and that combination of DM mass and cross section could not be excluded. 

\section{Results}\label{sec:Results}

No significant excess of events was observed above the expected background rate based on a background only fit to the data, so the procedure outlined above was used to calculate a set of bands of DM mass and DM-nucleon cross section that are excluded at the 90\% confidence level. 
This process was repeated for both inelastic scattering channels considered in this analysis.
The region outside the bands could not be excluded because more extreme cross sections result in too few signal events; small cross sections produce a small signal rate and large cross sections cause stronger attenuation of the signal.

\begin{figure}[htpb]
  \centering
  \includegraphics[width=\textwidth]{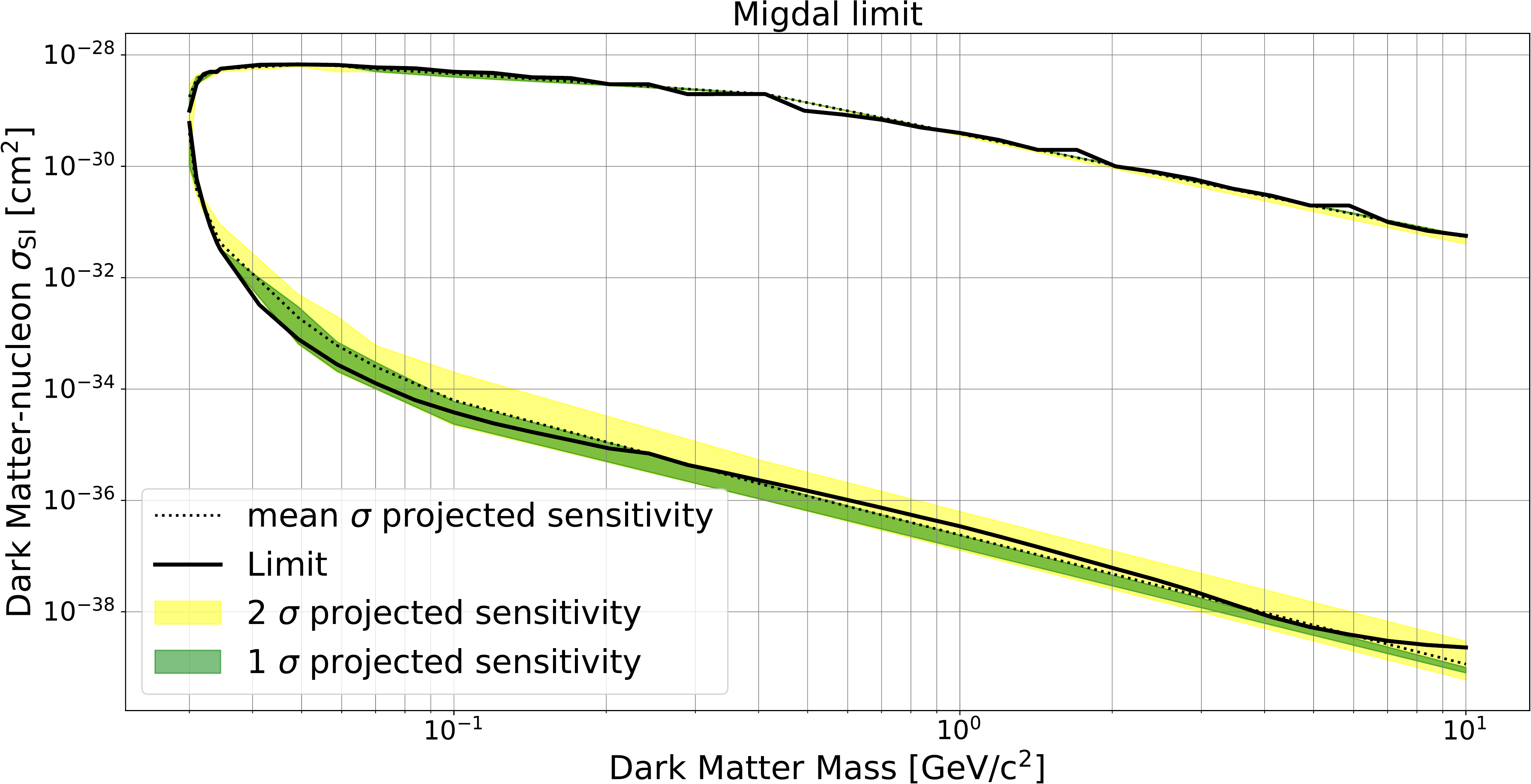}
  \caption{Mean (dotted line) expected sensitivity with 1$\sigma$ (green) and 2$\sigma$ (yellow) bands and the final limit (solid line). 
  }
  \label{fig:ProjSen}
\end{figure}

\begin{figure}[htpb]
    \centering
    \includegraphics[width=\textwidth]{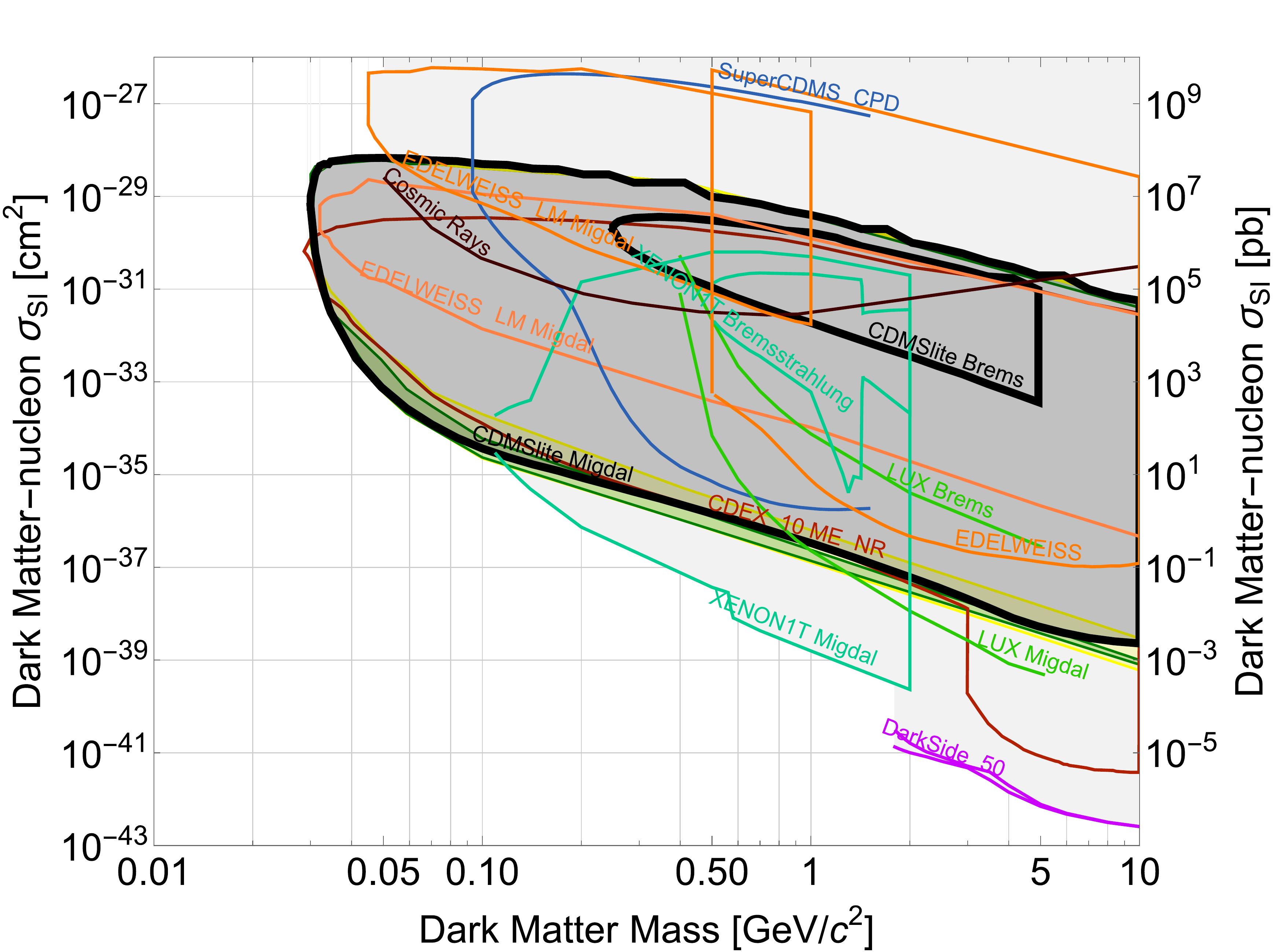}
    \caption{Current state of low-mass DM direct detection searches, with results from  
    SuperCDMS-CPD ~\cite{Alkhatib:2020slm}, DAMIC ~\cite{DAMIC:2020cut},
    cosmic ray bounds from Collar ~\cite{PhysRevD.98.023005}, and Darkside-50 ~\cite{Darkside_50} for 
    traditional elastic interaction searches. 
    The other curves are published limits from inelastic channel searches: LUX ~\cite{LUX_Inelastic}, EDELWEISS ~\cite{EDELWEISS_Migdal,PhysRevD.106.062004},
    XENON1T ~\cite{Aprile:2019jmx},
    CDEX ~\cite{CDEX_Inelastic},
    CDEX-10~\cite{PhysRevD.105.052005},  and this result.
    The green and yellow bands surrounding the Migdal result indicate the 1$\sigma$ and 2$\sigma$ projected sensitivity ranges, as shown in Figure \ref{fig:ProjSen}.
}
\label{fig:New_status_limits}
\end{figure}

Figure \ref{fig:ProjSen} shows the observed limit and projected sensitivity of the Migdal search estimated using pseudo datasets. 
\Cref{fig:New_status_limits} shows the exclusion regions and the current state of low-mass DM direct detection searches, including bremsstrahlung and Migdal channel searches.
New parameter space that was not previously tested by other DM searches is excluded via the Migdal channel for DM masses between 0.032 and 0.1 $\textrm{GeV}/c^2$. Although the bremsstrahlung result presented in this paper does not exclude any new parameter space, it is the most sensitive search using this channel for DM masses between 0.22 and 0.4 $\textrm{GeV}/c^2$. 
To test the sensitivity of the limit to yield modelling, we cross-checked our dependence by removing the Gaussian uncertainty constraint term in the likelihood and allowed our model to float unconstrained in the fit, corresponding to no prior knowledge of yield. 
The impact on the final limit was negligible.

The low CDMSlite energy threshold allows the bremsstrahlung channel limit to extend below the $0.4~\textrm{GeV}/c^2$ mass reach of the LUX and XENON1T bremsstrahlung analyses. However, both experiments have larger exposures, and so place lower cross section limits at higher masses~\cite{LUX_Inelastic,Aprile:2019jmx}.

Comparing the various Migdal channel results, this result also extends to lower masses than the LUX and XENON1T limits because of the lower threshold, and is competitive at intermediate masses despite a smaller exposure~\cite{LUX_Inelastic,Aprile:2019jmx}.
CDEX has an energy threshold that is 2--3 times higher than that of CDMSlite, thus the integrated rate of the Migdal signal is smaller and results in a slightly less sensitive limit~\cite{CDEX_Inelastic}. 
The EDELWEISS Migdal limit is not as sensitive in cross section due to the low exposure from operating a $33.4~\textrm{g}$ detector for $24~\textrm{hours}$. However, the shallow depth at which the EDELWEISS dataset was acquired allows the exclusion region to extend to higher cross sections than this analysis because these particles would lose their energy from scattering in the Earth before reaching the SuperCDMS experiment~\cite{EDELWEISS_Migdal}.
Similarly,  SuperCDMS-CPD data were collected at a surface facility using a low threshold detector searching for direct NR events~\cite{Alkhatib:2020slm}, making these data a prime candidate for repeating a bremsstrahlung or Migdal analysis in the future.

In summary, this analysis of CDMSlite data accounts for the shielding of strongly interacting DM particles by the Earth and atmosphere. This was implemented by calculating a damping parameter for the DM velocity distribution, and includes angular dependence of the incident DM. Using a profile likelihood framework, the bremsstrahlung channel was not found to probe any new parameter space, but the Migdal effect channel excludes new parameter space between 0.032 and 0.1 $\textrm{GeV}/c^2$.

\vspace{4mm}

\section{Acknowledgements}\label{Acknowledgement}

The authors would like to thank 
Matthew Dolan,
Timon Emken, 
Masahiro Ibe,
Felix Kahlhoefer,
Chris McCabe,
Wakutaka Nakano,
Jayden Newstead,
Yutaro Shoji, and
Kazumine Suzuki
for their useful discussions.

The SuperCDMS collaboration gratefully acknowledges technical assistance from the staff of the Soudan Underground Laboratory and the Minnesota Department of Natural Resources. The CDMSlite and iZIP detectors were fabricated in the Stanford Nanofabrication Facility, which is a member of the National Nanofabrication Infrastructure Network, sponsored and supported by the NSF. Funding and support were received from the National Science Foundation, the U.\ S.\ Department of Energy (DOE), Fermilab URA Visiting Scholar Grant No.\  15-S-33, NSERC Canada, the Canada First Excellence Research  Fund, the Arthur B.\  McDonald Institute (Canada),  the Department of Atomic Energy Government of India (DAE), the Department of Science and Technology (DST, India) and the DFG (Germany) - Project No.\ 420484612 and under Germany’s Excellence Strategy - EXC 2121 ``Quantum Universe" – 390833306.   Fermilab is operated by Fermi Research Alliance, LLC,  SLAC is operated by Stanford University, and PNNL is operated by the Battelle Memorial Institute for the U.S.\  Department of Energy under contracts DE-AC02-37407CH11359, DE-AC02-76SF00515, and DE-AC05-76RL01830, respectively.

\begin{table*}[htpb]
    \begin{tabular}{|l|c|c|c|c|c|c|}
    \hline
    & Atmosphere & Soudan & Earth Crust & Earth Mantle & Earth Outer Core & Earth Inner Core\\ 
    \hline
    Outer Radius [km] & \Cref{tab:atmos} & 0.7135 (depth) & 6.371 x $10^3$ & 6.331 x $10^3$ & 3.46 x $10^3$  & 1.22 x $10^3$\\ 
    \hline
    Density [g/cm$^3$] & \Cref{tab:atmos} & see aux data & 3.1 & 5.514 & 11 & 12.6\\ 
    \hline
    Mass Fraction H & &  & & & & \\
    Mass Fraction C & 0.0002 &  & & & & \\
    Mass Fraction O & 0.231 & & 0.476 & 0.503 & 0.0273 & 0.0273 \\
    Mass Fraction Na & &  & 0.029 & & & \\
    Mass Fraction Mg & &  & 0.015 & 0.256 & & \\
    Mass Fraction Al & &  & 0.083 & & & \\
    Mass Fraction Si & &  & 0.283 & 0.241 & 0.0509 & 0.0509 \\
    Mass Fraction P & &  & & & & \\
    Mass Fraction K & &  & 0.027 & & & \\
    Mass Fraction Ca & &  & 0.037 & & & \\
    Mass Fraction Mn & &  & & & & \\
    Mass Fraction Fe & &  & 0.051 & & 0.8509 & 0.8509 \\
    Mass Fraction Ti & &  & & & & \\
    Mass Fraction S & & & & & 0.0188 & 0.0188\\
    Mass Fraction N & 0.756 & & & & & \\
    Mass Fraction Ar & 0.013 & & & & & \\
    Mass Fraction Ne & 0.00001 & & & & & \\
    Mass Fraction Ni & & & & & 0.0520 & 0.0520 \\
    \hline
    \end{tabular}
    \caption{Model parameters used in the calculation of the velocity distribution damping.~\cite{2003TrGeo...3.....R,Damascus_paper,Damascus_code,moller2014chemistry,core_model}}
\label{tab:velodamp}
\end{table*}

\section{Appendix}\label{app:Soudan}

We describe the rock and chemical composition in \texttt{SoudanRegion.csv} and \texttt{RockChem.csv}, respectively. 
The \texttt{SoudanRegion.csv} file contains the area and fraction of each rock type in eight directions between radii of 100, 500, 1000, 5000, 10000, 20000 and 50000 meters.
The elemental mass fractions for the chemical composition is in \texttt{RockChem.csv}. 
Parameters for the earth and atmosphere are listed in Table \ref{tab:velodamp}. The density of the Earth was taken from Ref. \cite{DZIEWONSKI1981297}.

\clearpage
\bibliography{bib/other,bib/ASF,bib/damping,bib/SCDMS,bib/signals}

\end{document}